# An Objective Evaluation Metric for image fusion based on Del Operator

Ali A. Kiaei[1,*], Hassan Khotanlou[1], Mahdi Abbasi[1], Paniz Kiaei[2], Yasin Bhrouzi[3]

[1] Computer science department, Bu-Ali Sina University, Hamadan, Iran
[2] Computer science department, Alzahra University, Tehran, Iran
[3] Department of Mathematics, University of Birjand

*Abstract:* In this paper, a novel objective evaluation metric for image fusion is presented. Remarkable and attractive points of the proposed metric are that it has no parameter, the result is probability in the range of [0, 1] and it is free from illumination dependence. This metric is easy to implement and the result is computed in four steps: (1) Smoothing the images using Gaussian filter. (2) Transforming images to a vector field using Del operator. (3) Computing the normal distribution function ($\mu, \sigma$) for each corresponding pixel, and converting to the standard normal distribution function. (4) Computing the probability of being well-behaved fusion method as the result. To judge the quality of the proposed metric, it is compared to thirteen well-known non-reference objective evaluation metrics, where eight fusion methods are employed on seven experiments of multimodal medical images. The experimental results and statistical comparisons show that in contrast to the previously objective evaluation metrics the proposed one performs better in terms of both agreeing with human visual perception and evaluating fusion methods that are not performed at the same level.

Keywords: Image fusion method, Objective evaluation metric, Del operator, probability distribution function, cumulative distribution function

## 1. Introduction

Traditionally, the results of image fusion methods have been compared to each other using subjective evaluations, but, for an image, it may vary based on observer's background. However, because of the importance of evaluating image fusion methods objective metrics were introduced as an alternate that are consistent with human visual perception.

There are two classes of objective evaluation metrics for images. In the first class, the image is compared to a ground-truth. So, quantitative comparisons can be implemented between them. Some of metrics in this class are: cross-correlation (CC), difference entropy (DE), mean absolute error (MAE), mutual information (MI), peak signal to noise ratio (PSNR), quality index (QI), root mean square error (RMSE), and structural similarity (SSIM) index [Zhao et al. 2007].

Since usually there is no ground-truth fused image to compare with, the second class was created, where they evaluate the fused image in the absence of ground-truth. So, the fused image may either be evaluated by some known indexes or be compared to the source images using some non-reference metrics. Some known indexes are standard deviation (STD), Average Gradient (AG), Entropy (Entrp), and Edge Intensity (EI). Some of the non-reference objective metrics, on the other hand, are as follows:
- Objective Pixel-level Image Fusion metric ($Q_g$) [Xydeas et al. 2000(a)]: the edge information in each pixel plays an important role in this metric. Its operation assumes that the edge information is related to the visual information. So, weighting them makes acceptable measure for quantifying the image fusion performance.

---

[*] Corresponding author: Kiaei@basu.ac.ir

- Objective gradient based Image Fusion Performance Measure ($Q_{AB/F}$) [Xydeas et al. 2000(b)]: it computes the fusion performance by calculating transferred edge information from the sources images into the fused image.
- MI [Qu et al. 2001]: as the most commonly used objective metric, it uses the mutual information to find how the salient features are dispersed during the fusion process.
- Piella's quality index ($Q_E$) [Piella et al. 2003]: its measure is based on image quality index and locally calculates the salient information that contained during the fusion.
- Cvejic's universal quality index ($Q_c$) [Cvejic et al. 2005]: based on the Universal Image Quality Index, it computes the weights using the similarity between local blocks in the source and fused images.
- Mutual Information using Tsallis Entropy (MI-TE) [Cvejic et al. 2006]: using a generalization of Shannon entropy, named as Tsallis entropy, this metric computes the mutual information between the fused image and source images.
- Zhao's phase congruency ($Q_p$) [Zhao et al. 2007]: the measure is based on feature and is computed by phase congruency and its corresponding moments.
- Yang's local structural similarity ($Q_y$) [Yang et al. 2008]: it estimates the preserved information by calculating local structural similarity. The operations for different local regions evaluate this similarity.
- Normalized Mutual Information measure (NMI) [Hossny et al. 2008]: this method, at first, normalizes all the mutual information between fused image and source images. Then, the obtained same scaled ones are combined as mutual information between the fused and source images.
- Visual Information Fidelity for Fusion (VIFF) [Han et al. 2013]: using the visual information fidelity of different sub-bands, a multi-resolution metric is computed on blocks of images.

Table 1 summarizes advantages and disadvantages of these non-reference metrics.

Table 1: Some known non-reference objective metrics for image fusion.

| method | Advantages | Disadvantages |
|---|---|---|
| $Q_{AB/F}$ [Xydeas et al. 2000(b)] | - Demonstrate the quality of visual information<br>- Extract important information that exist in source and fused image<br>- Measure the ability of transferring information | - NOT clear that a significance feature is related to what level of edge strength |
| MI [Qu et al. 2001] | - Agree well with informal subjective tests | - Tend to ranked best the averaging fusion method [Cvejic et al. 2006]<br>- Without a ground-truth, it does NOT make sense [Zhao et al. 2007].<br>- performing fusion in different levels leads incorrect result [Yang et al. 2008].<br>- Mixing un-normalized mutual information leads false result [Hossny et al. 2008]. |
| $Q_E$ [Piella et al. 2003] | - Locally, Compare the large saliency of source images and the fused one.<br>- The weights are correlated to the saliency of the source images.<br>- The measure is in the range of [0, 1].<br>- Increasing α grows the importance of illumination in out-of-focus images.<br>- The measure is correlated to the edge of images. | - Different parameters such as window size, α, and weights change the result.<br>- The weights do NOT measure the similarity [Zhao et al. 2007].<br>- NOT good when conflicting or complementary source images were fused [Yang et al. 2008] |
| $Q_c$ [Cvejic et al. 2005] | - Specialize MSE and MI for image fusion methods | - Saliency makes no sense.<br>- Redundant features make the metric invalid [Zhao et al. 2007]. |
| MI-TE [Cvejic et al. 2006] | - Having no parameter | |
| $Q_p$ | - Count how the features are fused | |
| [Zhao et al. 2007] | - Differentiate the performance of fusion methods<br>- Variety properties of images are evaluated in this metric. | |
| $Q_y$<br>[Yang et al. 2008] | - Manage redundant regions<br>- Manage complementary and conflicting regions<br>- Manage different fusion levels | |
| NMI<br>[Hossny et al. 2008] | - Estimate the transferred information between fused and source images<br>- Differences between entropies don't affect the result | |
| VIFF<br>[Han et al. 2013] | - Lower time complexity than other fusion metrics<br>- Lower computational complexity than other metrics<br>- Improved in predictive performance | - Only gives an approximate estimation of fusion methods' performance. |

## 2. Background

In this section, the requirement mathematical backgrounds for proposed objective evaluation are reviewed briefly. At first, Del operator is defined and it is shown that how an image is transformed from spatial space to the vector one, using Del operator. Then, a summary of cumulative distribution function of the normal distribution is considered, which is used in the proposed objective evaluation metric.

### Transforming images using Del operator

As a vector differential operator, Del is widely used in mathematics. Standard derivative and gradient are two important topics in calculus that are defined when it applied to a 1D function and a field, respectively. Using partial derivative operators in $R^n$ with coordinates $(x_1, \ldots, x_n)$, Del is defined by:

$$\nabla = \left(\frac{\partial}{\partial x_1}, \ldots, \frac{\partial}{\partial x_n}\right) = \sum_{i=1}^{n} \frac{\partial}{\partial x_i} \vec{e_i} \tag{1}$$

By assuming $D$ as a rectangle territory, $F = F(x,y) = P(x,y)i + Q(x,y)j$ can be the Gradient field on $D$. again, if $\nabla$ indicates gradient, there is a U(x,y) on $D$ that $F = \nabla U$, and then $F$ is conservative [Mercer 2014] :

$$P = \frac{\partial U(x,y)}{\partial x}, \quad Q = \frac{\partial U(x,y)}{\partial y} \tag{2}$$

Finite differences are acceptable approximation of derivatives in discrete cases. There are three types of finite differences: forward, backward, and central direction. Because of its advantages, the central one is used to transform the images using Del operator.

Like the continuous mode, the existence of such $U$ that $D_U = P_U i + Q_U j$ can be proved in discrete case, where $P_U$ and $Q_U$ are defined as follows:

$$P_U = \frac{\partial U(x,y)}{\partial x} = \lim_{h \to 0} \frac{U(x+h,y) - U(x-h,y)}{2h} \tag{3}$$

$$Q_U = \frac{\partial U(x,y)}{\partial y} = \lim_{k \to 0} \frac{U(x,y+k) - U(x,y-k)}{2k} \tag{4}$$

where $h, k \to 0$. In digital images the nearest neighbor of a pixel has a distance of pixel width. In other words, the smallest possible values for $h, k$ in digital images are 1. The linear transformation conserves the additivity and homogeneity conditions. So, the conditions are also satisfied in central direction. By assuming $h, k=1$, the above equations become:

$$P_U(x,y) = \big(U(x+1,y) - U(x-1,y)\big) / 2 \tag{5}$$

$$Q_U(x, y) = (U(x, y + 1) - U(x, y - 1)) / 2 \qquad (6)$$

By implementing the above equations for each pixel independently, images can be transformed to a vector space domain. In other words, each pixel is converted to a vector that its decompositions in the Cartesian coordinate system along the $\hat{x}$ and $\hat{y}$ directions are $P_U$ and $Q_U$ respectively.

### Cumulative distribution function

The z table is standard normal table that includes values of cumulative distribution function ($\Phi$) of the normal distribution. This function computes the probability of a range of statistic observation. As there are infinite normal distributions, one of the simplest ways to find the probability of a range is by converting the normal distribution to the standard one and then checking the standard normal table. The standard normal distribution has the mean of zero ($\mu = 0$), and standard derivative of one ($\sigma = 0$). So, converting x from a normal distribution with $\mu, \sigma$ to the standard normal distribution is as follows:

$$z = \frac{x-\mu}{\sigma} \qquad (7)$$

where $z$ is called the z-score of $x$. Finding the probability of $z$ from the cumulative table means the probability that a statistic is less than $z$ $(P(-\infty < statistic < z))$. However, the values of this table are computed by:

$$\Phi(z) = \frac{1}{\sqrt{2\pi}} \int_{-\infty}^{z} e^{-t^2/2} dt \qquad (8)$$

As an alternative function, the complementary cumulative function means the probability of being greater than z:

$$f(z) = 1 - \Phi(z) \qquad (9)$$

## 3. Objective Evaluation metric using Del operator ($Q_{Del}$)

Integrating complementary information of two or more images is the aim of image fusion. So, an objective evaluation metric should estimate how much the important features are transferred from the source images to the fused one.

The proposed objective evaluation metric uses cumulative distribution function in the new domain, to evaluate the image fusion performance as a probability.

To implement the proposed objective evaluation metric, consider an image fusion method is employed on $n$ source images $(I_1, \ldots, I_n)$ to fuses them as $I_F$. As shown in Fig. 1, to reduce the negative effect of noise, at first, all source images and fused one are smoothed using a two-dimensional Gaussian function $G_\sigma(x, y)$ with standard deviation $\sigma$, as follows:

$$I(x, y) = G_\sigma(x, y) * I(x, y) \qquad (10)$$

Using finite differences in equations 5-6, for each pixel $(x,y)$ from smoothed version of source images and the fused one, the proposed objective evaluation metric computes $P(x,y)$ and $Q(x,y)$. For smoothed image $I$, the transformed image $mI$ is computed by:

$$mI(x, y) = P_I(x, y)^2 + Q_I(x, y)^2 \qquad (11)$$

Eq. 11 is implemented on all smoothed source images and the fused one. So, the transformed source images are $(mI_1, \ldots, mI_n)$, and the transformed fused image is $mI_F$. From the transferred source images and for all $(x, y)$, the proposed method defines $\mu(x, y), \sigma(x, y)$ as follows:

$$\mu(x,y) = \frac{mI_1^2(x,y)+\cdots+ mI_n^2(x,y)}{mI_1(x,y)+\cdots+mI_n(x,y)} \quad (12)$$

$$\sigma(x,y) = \sqrt{\frac{\sum_{i=1}^{n}(mI-\mu)^2}{n}} \quad (13)$$

Now, the fused image's z-score at each pixel $(x,y)$ is calculated by:

$$z(x,y) = \left|\frac{mI_F(x,y) - \mu(x,y)}{\sigma(x,y)}\right| \quad (14)$$

By calculating $f(z(x,y))$ from eq. 9 (or from the complementary cumulative table), the probability of fusing well in each pixel is computed as follows:

$$P(x,y) = 2f(z(x,y)) \quad (15)$$

So, $P(x,y)$ finds the probability that a statistic has a more distance to the zero than $z(x,y)$. In other words, at point $(x,y)$, with what probability a well fused pixel can be find that $mI_F(x,y)$ is better than its $mI(x,y)$, regarding normalization. The values of $z(x,y)$, $f(z(x,y))$ and $P(x,y)$ are shown in Fig. 2.

Finally, the proposed objective evaluation metric for those $n$ source images $(I_1, \ldots, I_n)$ and the fused image $(I_F)$ is as follows:

$$Q_{Del}(I_1, \ldots, I_n; I_F) = \frac{\sum_x \sum_y (P(x,y))}{\sum_x \sum_y 1} \quad (16)$$

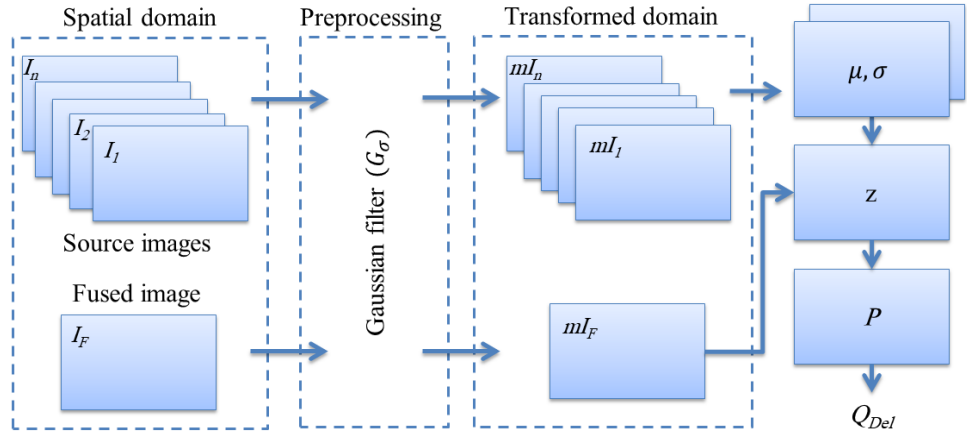

Fig. 1: A schematic of proposed objective evaluation metric using Del operator (the metric is available online as $Q_{Del}$ at RIV lab[1]).

---

[1] http://riv.basu.ac.ir/Research/ImageFusion/ObjectiveEvaluations.aspx

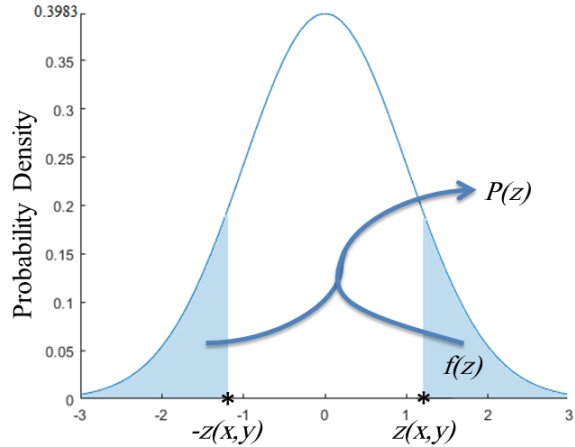

Fig. 2: plot of the standard probability density functions together with showing $z(x,y), f(z(x,y))$ and $P(x,y)$.

$Q_{Del}$ is sensible to the saliency and edges, where it is match to human visual system [Xydeas et al. 2000b].

## 4. Statistical comparison

Different MR and CT images have been served in this section to make some experiments; showing how the proposed metric works and what advantages it has compared to the previous objective evaluation methods. The images are 256x256 and can be downloaded from Aanlib[1]. As shown in Fig. 3, the aim in experiments is using objective evaluations for each two images with complementary information that are fused. The experiments demonstrate patients with acute stroke, cerebral toxoplasmosis, vascular dementia, and AIDS dementia.

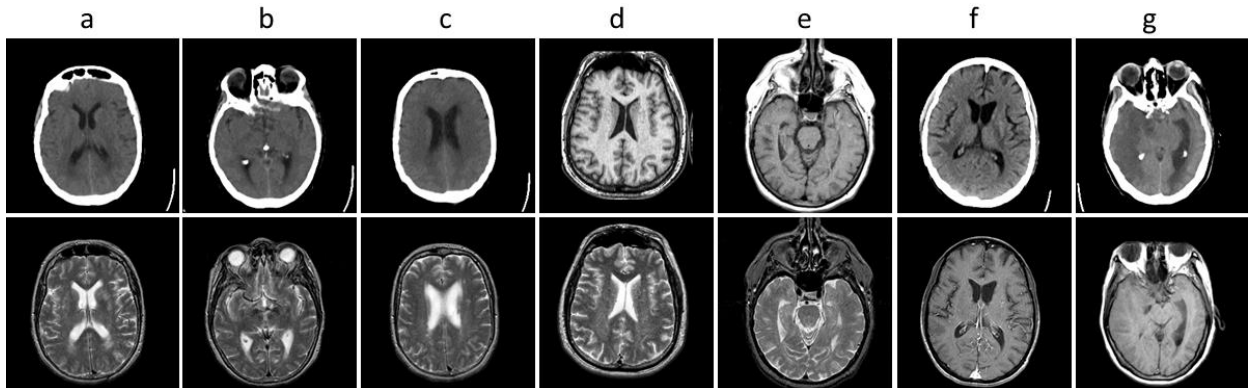

Fig. 3: different complementary source images. (a) Experiment 1: CT and MR-T2 images of brain with acute stroke; (b) Experiment 2: CT and MR-T2 images of brain with acute stroke; (c) Experiment 3: The transaxial MR-T1 and MR-T2 images of the normal brain; (d) Experiment 4: MR-T1 and MR-T2 images of the brain of patients with vascular dementia; (e) Experiment 5: CT and MR-T2 images of the brain of patients with vascular dementia; (f) Experiment 6: CT and MR-GAD images of the brain of patients with vascular dementia; (g) Experiment 7: CT and MR-T1 images of the brain of patients with vascular dementia.

The proposed objective evaluation metric was compared to some recently and known previous metrics on eight known fusion methods to verify its advantages. The comparing fusion methods are as follows: the discrete wavelets transform (DWT), FSD Pyramid (FSD), Gradient Pyramid (GP), Laplacian Pyramid (LP), Morphological Difference Pyramid (MDP), PCA, Ratio Pyramid (RP), and Shift-Invariant Discrete Wavelet Transformation with Harr wavelet (SIDWT). Their objective evaluation results have been shown in comparison with the proposed metric (fusion methods are available at RIV Lab[1]).

---

[1] http://www.med.harvard.edu/aanlib/home.html

The comparing objective evaluation metrics for image fusions in this paper are: STD, Entropy, EI, SSIM, Xydeas's metric [2000(a)], $Q_{AB/F}$ [Xydeas et al. 2000(b)], MI [Qu et al. 2001], Piella's metric [2003], Cvejic's metric [2005], Zheng's metric [2007], Zhao's metric [2007], Chen's metric [2007] and Wang's metric [2008].

Fig. 4 demonstrated the results of different previously known objective evaluation metrics on some image fusion methods, for all experiments in Fig. 3 separately. They were evaluated by previously metrics and the average rank of each fusion method was shown in this figure. For example in the 1st experiment, LP and DWT were the best and worst average ranked fusion methods, when all 13 objective evaluation metrics were considered. As the same way, other lines demonstrated similar results, ranks and average ranks on experiments 2-7.

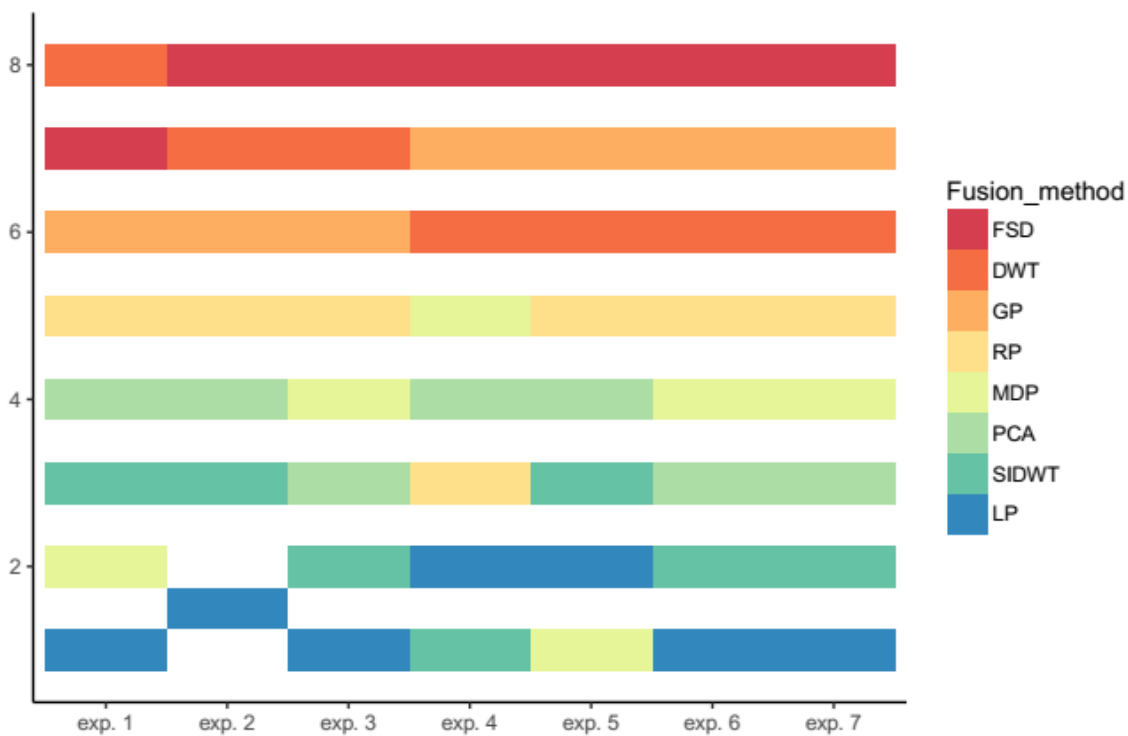

Fig. 4: ranks of fusion methods on all Experiment in Fig. 3.

In Table 2, on the other hand, the results of using the proposed method as an objective evaluation metric were shown. The values are between [0,1], where the bigger values indicate better fusion method. Additionally, ranks of fusion methods on each experiment were demonstrated in parentheses.

In table 3, the correlation coefficient between each metric's ranks and the average of the others' ranks were computed, experiment by experiment. As before, the ranks of these correlations on metrics were calculated in the table and were shown in parentheses. As it can be shown in this table, the ranks of proposed metric, $Q_{Del}$, were most correlated to the average ranking of other metrics in experiments 2, 3 and 7 with 96, 95 and 96 percent, respectively. Although it were ranked two in experiments 1, 2 and ranked three in experiments 5, 6, the last column shows that $Q_{Del}$ had the best average rank when all the experiments are collectively considered. Based on those 7 experiments in Fig. 3, the general ranking of objective evaluation metrics were shown in Table 4. As shown in this table, the proposed objective evaluation metric, $Q_{Del}$, had the best rank among the metrics, when all experiments are considered.

---
[1] http://riv.basu.ac.ir/Research/ImageFusion/FusionMethods.aspx

Table 2: proposed metric results ($Q_{Del}$) and using it to rank fusion methods on all 7 Experiments in Fig. 3.

| $Q_{Del}$ | DWT | FSD | GP | LP | MDP | PCA | RP | SIDWT |
|---|---|---|---|---|---|---|---|---|
| exp.1 | 0.41(6) | 0.3(7) | 0.3(8) | 0.68(1) | 0.68(2) | 0.55(5) | 0.58(4) | 0.67(3) |
| exp.2 | 0.41(6) | 0.28(8) | 0.28(7) | 0.64(2) | 0.65(1) | 0.5(5) | 0.53(4) | 0.64(3) |
| exp.3 | 0.42(6) | 0.31(7) | 0.31(8) | 0.69(1) | 0.67(3) | 0.58(5) | 0.61(4) | 0.68(2) |
| exp.4 | 0.35(6) | 0.21(8) | 0.21(7) | 0.51(4) | 0.51(2) | 0.48(5) | 0.51(3) | 0.54(1) |
| exp.5 | 0.32(4) | 0.23(8) | 0.23(7) | 0.41(3) | 0.52(1) | 0.28(6) | 0.29(5) | 0.48(2) |
| exp.6 | 0.38(6) | 0.26(7) | 0.26(8) | 0.62(2) | 0.59(3) | 0.58(4) | 0.56(5) | 0.63(1) |
| exp.7 | 0.39(6) | 0.26(8) | 0.26(7) | 0.63(1) | 0.57(3) | 0.54(5) | 0.56(4) | 0.62(2) |

Table 3: the correlation coefficient between each metric and average result of other ones. The rank of each metric is computed in the parentheses, experiment by experiment.

| Correlation with whole | exp.1 | exp.2 | exp.3 | exp.4 | exp.5 | exp.6 | exp.7 | average rank |
|---|---|---|---|---|---|---|---|---|
| STD | 0.46(9) | 0.63(7) | 0.41(8) | 0.62(8) | 0.68(7) | 0.56(8) | 0.28(9) | 8 |
| Entrp | 0.42(10) | 0.52(9) | 0.04(12) | 0.78(5) | 0.98(1) | 0.65(6) | 0.66(6) | 7 |
| EI | 0.33(11) | 0.36(11) | 0.13(10) | 0.21(10) | 0.41(10) | -0.03(12) | 0.07(11) | 10.71 |
| SSIM | 0.86(4) | 0.85(4) | 0.93(2) | 0.87(3) | 0.79(5) | 0.85(5) | 0.8(5) | 4 |
| MI2 | 0.69(5) | 0.71(5) | 0.59(6) | 0.64(7) | 0.61(8) | 0.55(9) | 0.57(7) | 6.71 |
| Qabf | 0.5(7) | 0.44(10) | 0.44(7) | 0.08(11) | 0.02(12) | -0.07(13) | -0.02(13) | 10.43 |
| Wang | 0.24(13) | -0.05(13) | -0.27(13) | -0.17(14) | 0.33(11) | 0.6(7) | -0.1(14) | 12.14 |
| Xydeas | 0.92(3) | 0.93(2) | 0.9(4) | 0.94(1) | 0.85(2) | 0.96(1) | 0.94(2) | 2.14 |
| Zheng | 0.49(8) | 0.62(8) | 0.35(9) | 0.32(9) | 0.46(9) | 0.15(11) | 0.29(8) | 8.86 |
| Zhao | 0.3(12) | 0.08(12) | 0.07(11) | 0.03(12) | -0.13(13) | 0.29(10) | 0.05(12) | 11.71 |
| Piella | 0.95(1) | 0.93(3) | 0.91(3) | 0.86(4) | 0.84(4) | 0.95(2) | 0.91(3) | 2.86 |
| Cvejic | 0.66(6) | 0.71(6) | 0.72(5) | 0.76(6) | 0.77(6) | 0.9(4) | 0.85(4) | 5.29 |
| Chen | -0.75(14) | -0.81(14) | -0.71(14) | -0.04(13) | -0.3(14) | -0.74(14) | 0.18(10) | 13.29 |
| $Q_{Del}$ | 0.94(2) | 0.96(1) | 0.95(1) | 0.88(2) | 0.85(3) | 0.95(3) | 0.96(1) | 1.86 |

Table 4: ranking of objective evaluation metrics based on their correlation coefficients with other methods' average results.

| method | STD | Entrp | EI | SSIM | MI2 | Qabf | Wang | Xydeas | Zheng | Zhao | Piella | Cvejic | Chen | $Q_{Del}$ |
|---|---|---|---|---|---|---|---|---|---|---|---|---|---|---|
| final rank | 8 | 7 | 11 | 4 | 6 | 10 | 13 | 2 | 9 | 12 | 3 | 5 | 14 | 1 |

## 5. Multivariate analysis (MVA)

To analyze a multivariate data set, MVA refers some techniques. Three most useful of them are Correlation analysis, Principal Component Analysis (PCA) and Cluster Analysis. The first one focuses on the relationship between variables. The second one summarizes the most important information of the multiple variables of the dataset and visualizes them. Finally, the third one identifies groups of observations that have similar profiles, based on the features.

In this section, some methods are provided to visualize multivariate the dataset containing fusion methods and objective evaluation metrics. One way to determine if there is a linear correlation between multiple variables is using scatter plot matrices.

In Fig. 5, the scatter plot of the Experiment 1 is shown. The objective evaluation metrics are written in top and right lines. Each of them is plotted against other ones in the boxes on the lower left hand side of the whole scatterplot (below the diagonal line). On the other hand, the boxes on the upper right hand side of the whole scatterplot are correlations of the plots on the lower left hand. For example, the left square in the second row is an individual scatterplot of $Q_{Del}$ and Xydeas metrics, with $Q_{Del}$ as the X-axis and Xydeas as the Y-axis, where their correlation is 0.941.

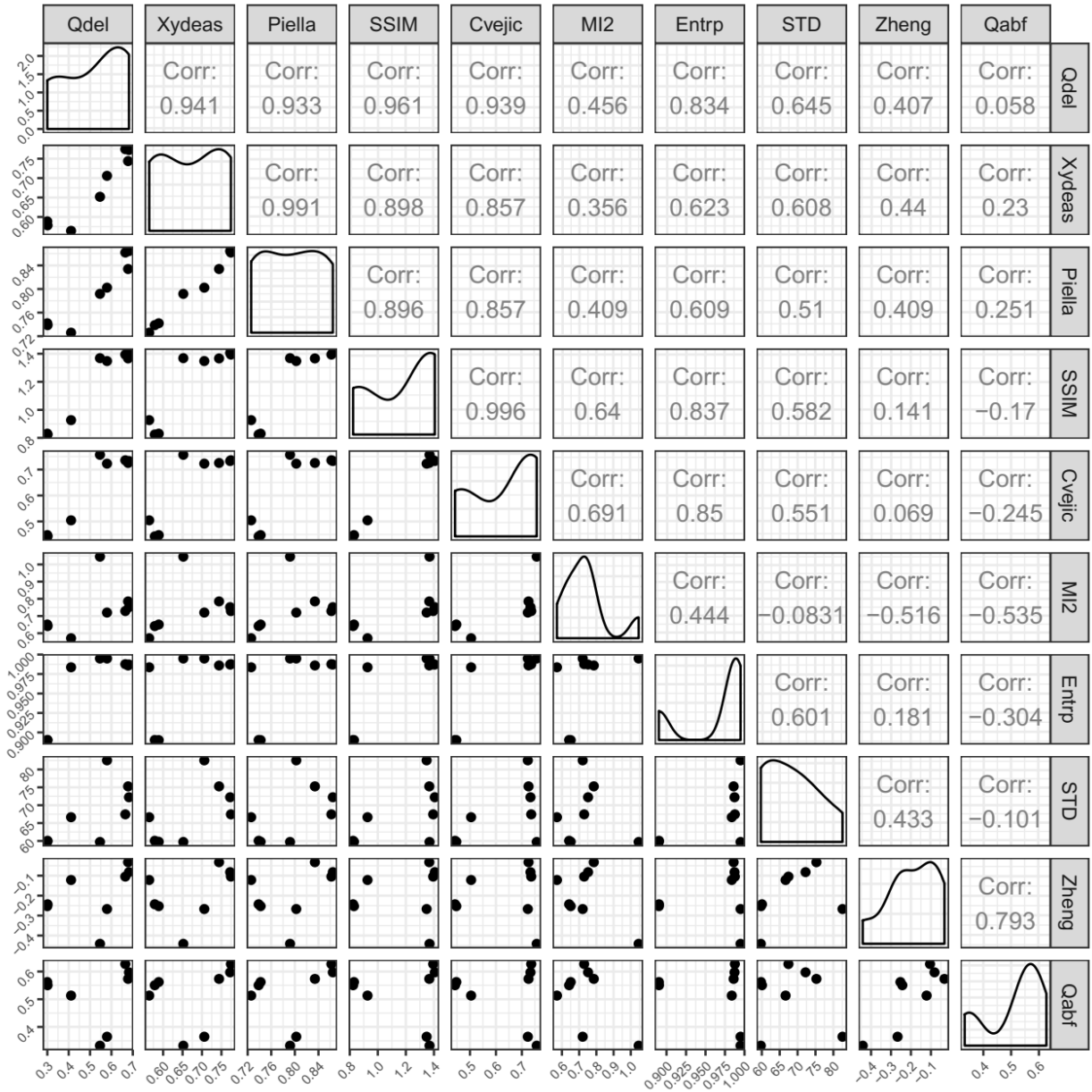

Fig. 5: a scatter plot matrix on Experiment 1. The plot contains the Scatter plot and the correlation coefficient between each pair of Objective Evaluation metrics, and Density distribution of each metric.

Fig. 6, shows an alternative scatter plot for experiment 2. The diagonal boxes in this figure show the histogram and distribution of each metric. The bottom left side of figure contains bivariate scatter plots and fitted lines on them. The values of correlations are shown in the top right boxes of figure. Moreover, their significance levels are starred as follows: symbols ("***", "**", "*", ".", " " ) mean p-values in the range of (0, 0.001, 0.01, 0.05, 0.1, 1), respectively. As shown in this figure, the most correlated metrics with $Q_{Del}$ are Xydeas, Piella, and SSIM. From the Xydeas, it is inferred that not only $Q_{Del}$ demonstrate the quality of visual information, but also it extracts important information that exist in source and fused image. On the other hand, Piella says that $Q_{Del}$ puts emphasis on the saliency of the source images.

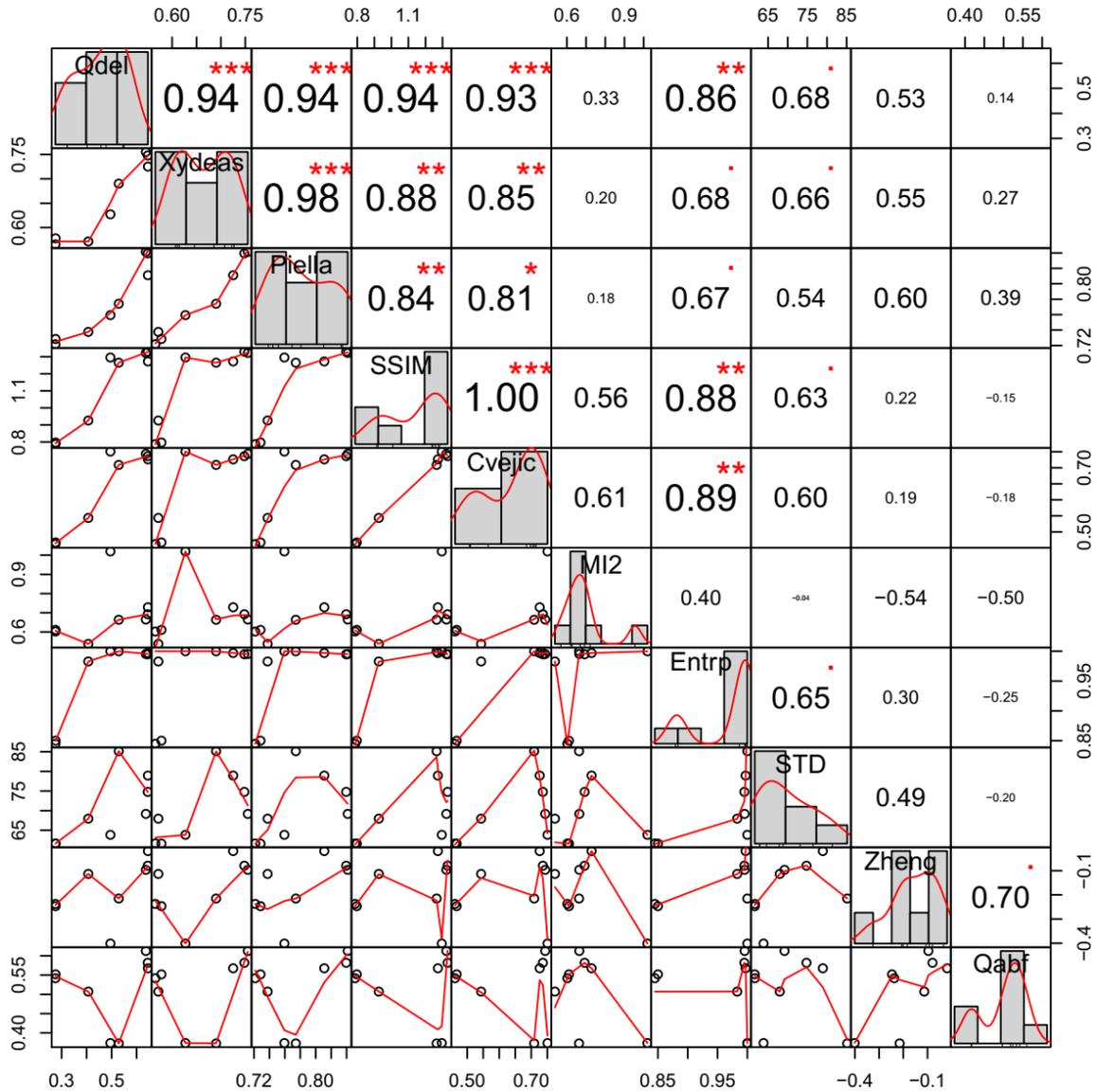

Fig. 6: Experiment 2's correlation coefficient and the significance levels as stars.

## Correlation analysis

As the first analysis for statistical evaluation for multivariate dataset, correlation analysis studies on the strength of relationship between each two variables that are numerically measured and continuous. One of the most common used is Pearson's correlation coefficient, where ranges from -1 as the (strongest negative correlation) to +1 (the strongest positive correlation possible).

In Fig. 7 a graphical display of a correlation matrix for Experiment 3 is shown. In this figure, the values of correlation coefficients are shown by color. Besides the colors, the sizes of circles are related to the degree of association between variables. In other words, the dark blue big circle shows the strong positive correlation between two variables. On the contrary, the transparent small circles show uncorrelated variables. The right side of the figure shows the legend color and corresponding correlation coefficient. However, p-value<0.01 is considered for the significance level, where insignificants values left blank.

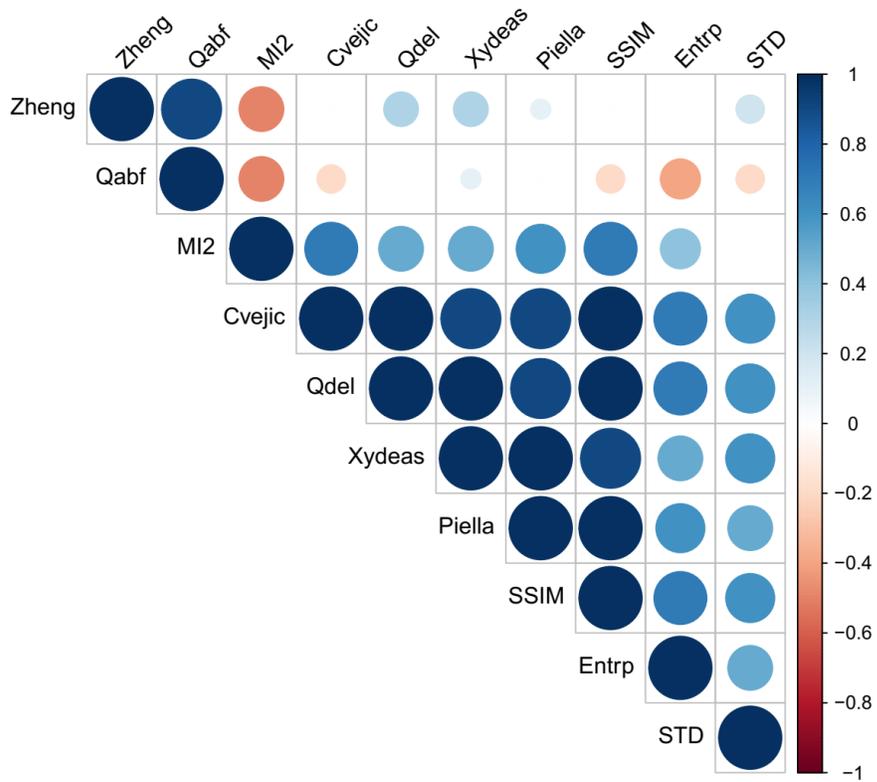

Fig. 7: The correlation matrix of Experiment 3.

Fig.8 recorded the correlation matrix of Experiment 4. Insignificant correlations are crossed. Besides the Xydeas and SSIM, in this experiment, one of the metrics that are highly correlated with $Q_{Del}$ is Cvejic. This means that the proposed metric specialize MSE and MI for image fusion methods.

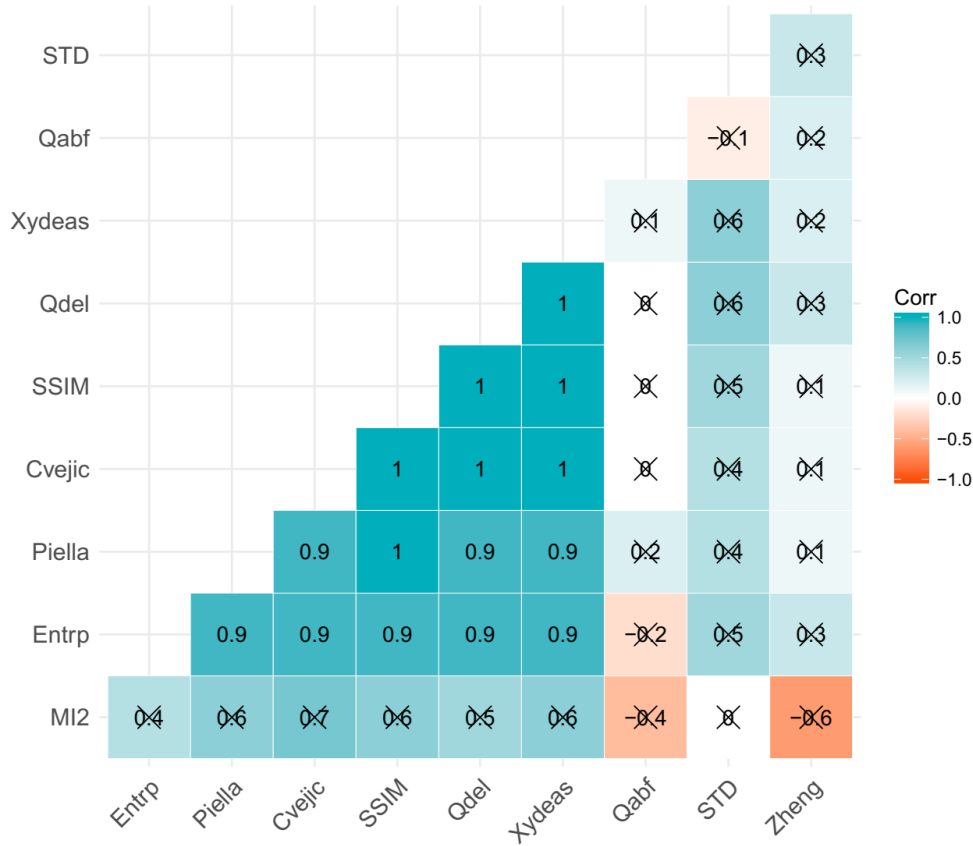

Fig. 8: the correlation matrix of Experiment 4 is recorded. The no significant coefficients are crossed.

### Principal component analysis (PCA)

To summarize and visualize the most important information in the dataset, PCA combine variables and reduces the dimension of features. Two most important dimensions in data can be a place of visualizing data, using a scatter plot.

Fig. 9 shows the PCA of Experiment 5. In this figure, dimensions (Dim1 and Dim2) reserve about 78% (56.6% + 21.7%) of the whole dataset's information. Metrics that are on the same (/opposite) side of the plot are with positive (/negative) correlations.

The figure shows $Q_{Del}$ use the information of Dim1 better than the Xydeas, Piella and SSIM, which were had the most correlated to the $Q_{Del}$ in the previous experiments. On the other hand, the projections of the $Q_{Del}$, Zheng, MI and Cvejic metrics on the diagonals of the coordinate axes have the most magnitudes in comparison with the other metrics. This projection is valuable since the axes are orthogonal and are independent from each other, and so, these metrics use the information of both coordinates more than the others.

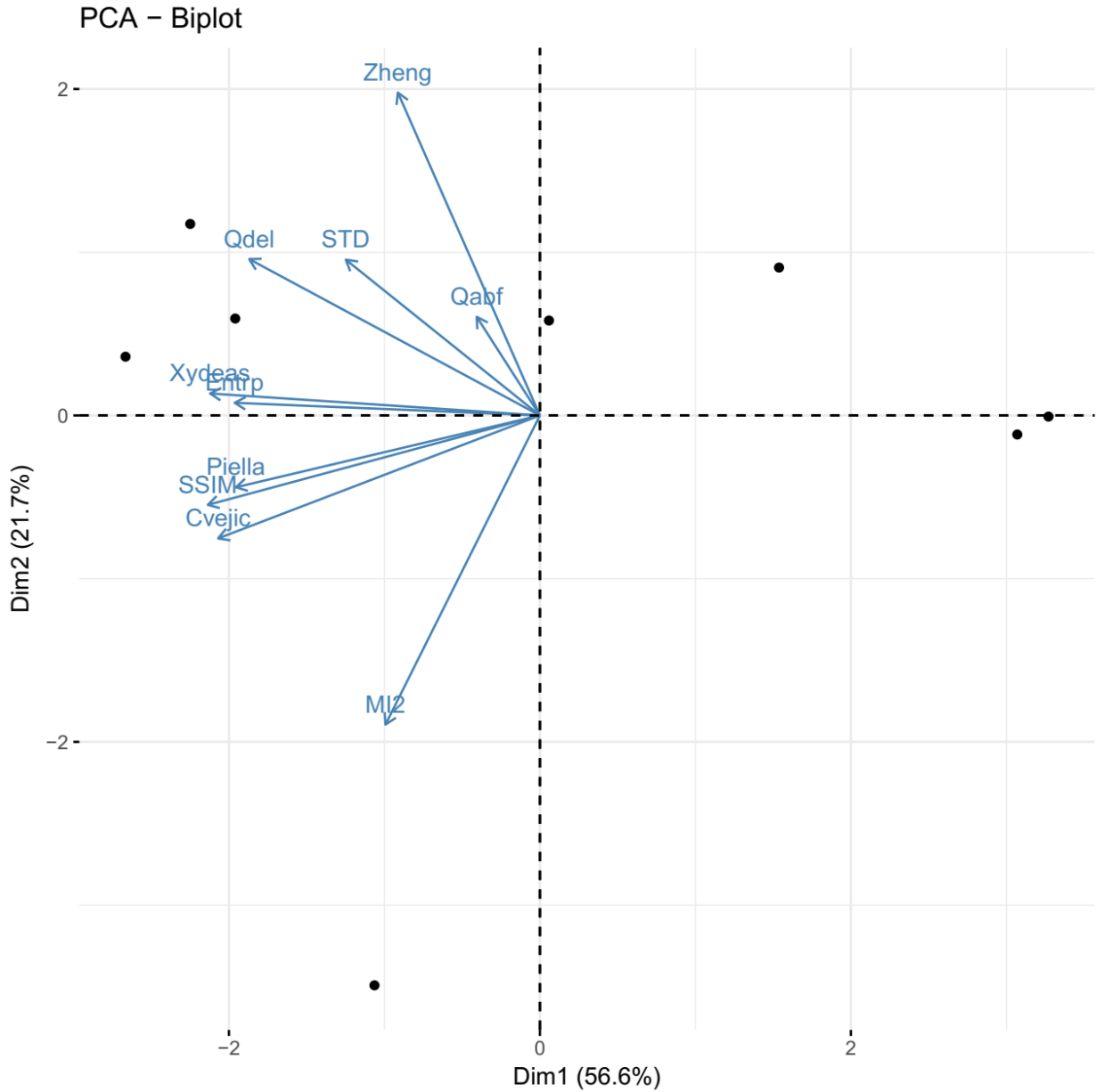

Fig. 9: PCA of Experiment 5. $Q_{Del}$, Zheng, MI2 and Cevjie get the information more than the other metrics.

### Cluster analysis

Another important study on metrics is cluster analysis, which categorizes groups of similar metrics contained by data. The similar metrics are pictured by hierarchical clustering as a tree named dendrogram. However the data may be in different scales that make them incomparable. To overcome the problem, the data should be normalized and then the cluster analysis is implemented.

In Fig. 10, a heatmap of Experiment 6 is shown. In this figure, the values are represented as colors. The hierarchical clustering and the dendrogram show the similarities.

As shown in this figure, the proposed $Q_{Del}$ metric is in the lowest leaf of the tree (the most similarity part), which means that the proposed $Q_{Del}$ metric use important information that almost five other metrics have similar results to it.

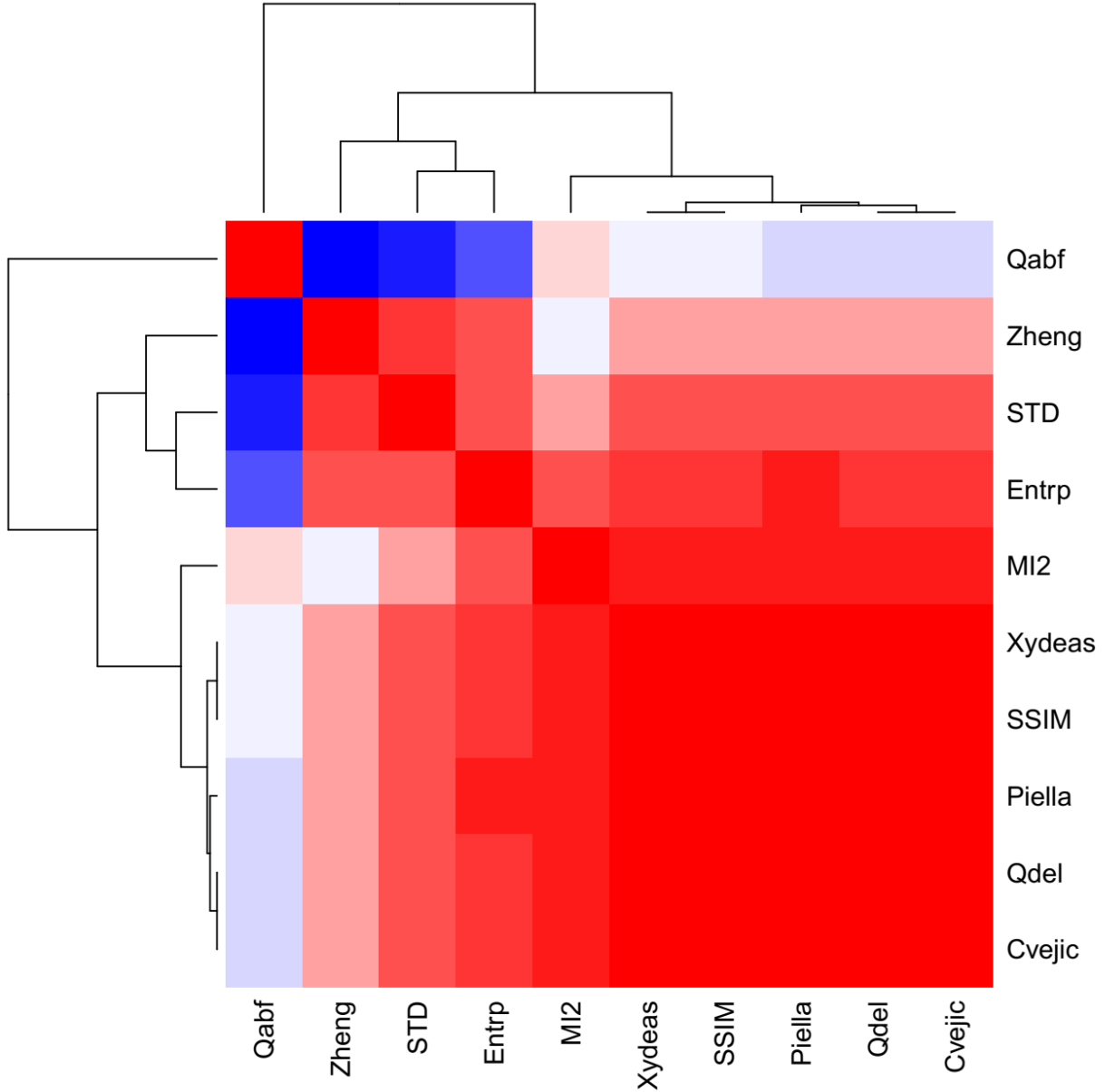

Fig. 10: heatmap of Experiment 6. The dendrogram of similar metrics shows that the proposed $Q_{Del}$ metric is in the most similarity part of the tree.

Besides the cluster analysis, two types of p-values are valuable: Approximately Unbiased (AU) p-value and Bootstrap Probability (BP) value. The first one measures the uncertainty using the p-vlaue of each cluster via multiscale bootstrap resampling. It tests the general hypothesis when the test bias is reduced, so that computes the confidence set of trees. On the other hand, the second one is computed by normal bootstrap resampling, which is more biased.

Fig 11 is an enhanced visualization of dendrogram for Experiment 7, which draws phylogenic trees. We assume that the number of groups for cutting the tree is 6. The AU p-values (%) are written in red, and the BP values are green. The highlighted red rectangles show the clusters with AU p-value > 0.95%. In other words, the hypothesis of H0 (the cluster does not exist) is rejected with significance level 0.05. The cluster rectangles show the similarity of proposed $Q_{Del}$ metric with some other metrics. Not only does $Q_{Del}$ overcome their disadvantages, but also the rectangles show that $Q_{Del}$ behave like them and so has their advantages.

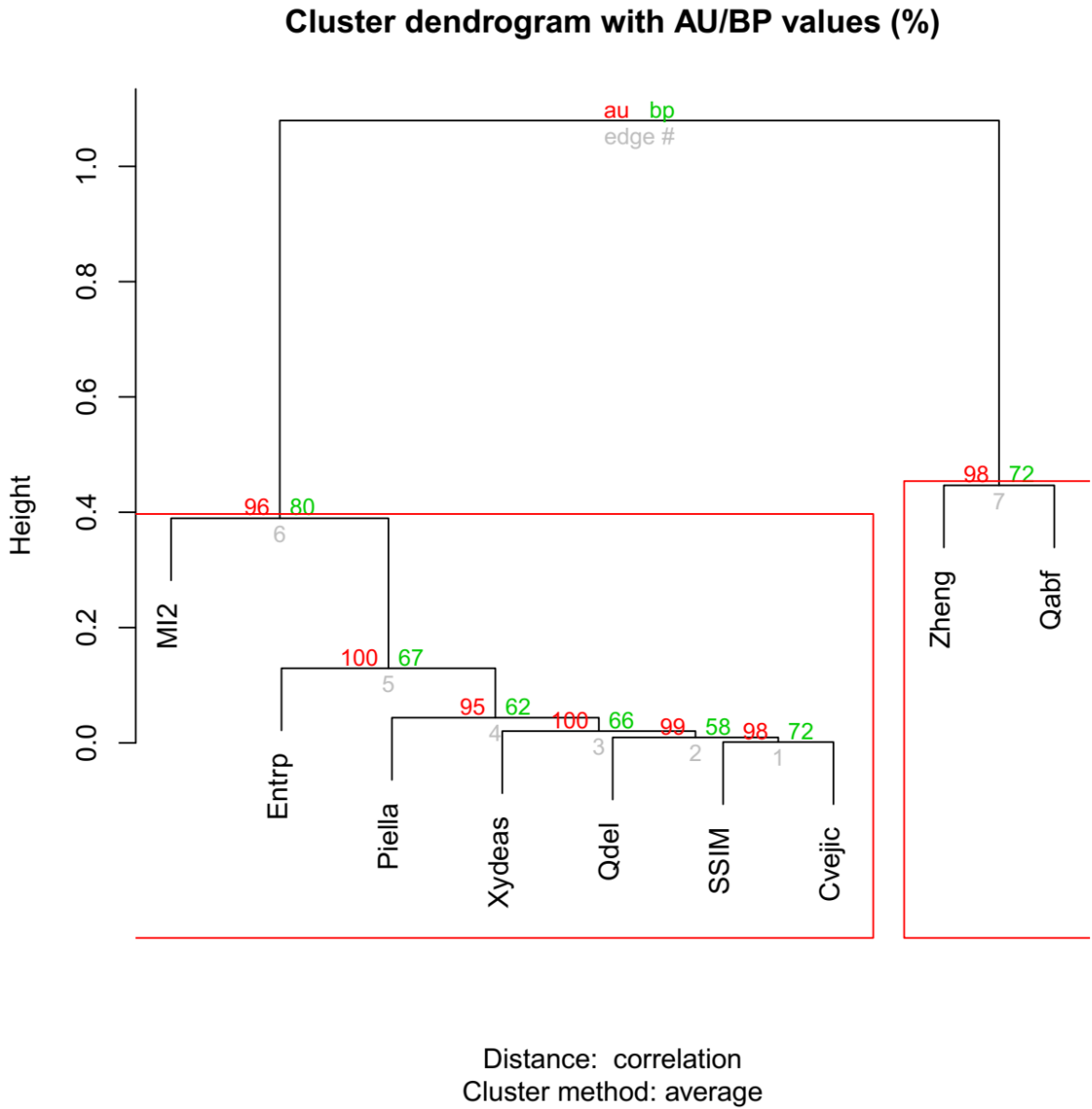

Fig. 11: the Cluster Dendogram of Experiment 7 with AU/BP Values.

As an alternative to heatmap, a more control over dimensions and appearance in Experiment 8 is shown in Fig. 12. The default clustering metric is Euclidean distance (instead of using Pearson correlation in the previous using heatmap). Moreover, the grouping variables are both objective evaluation metrics and fusion methods. Based on hierarchical clustering, the metrics are divided into three clusters. The hierarchical clustering separate Entropy, SSIM, and Zheng metrics from the $Q_{Del}$'s group.

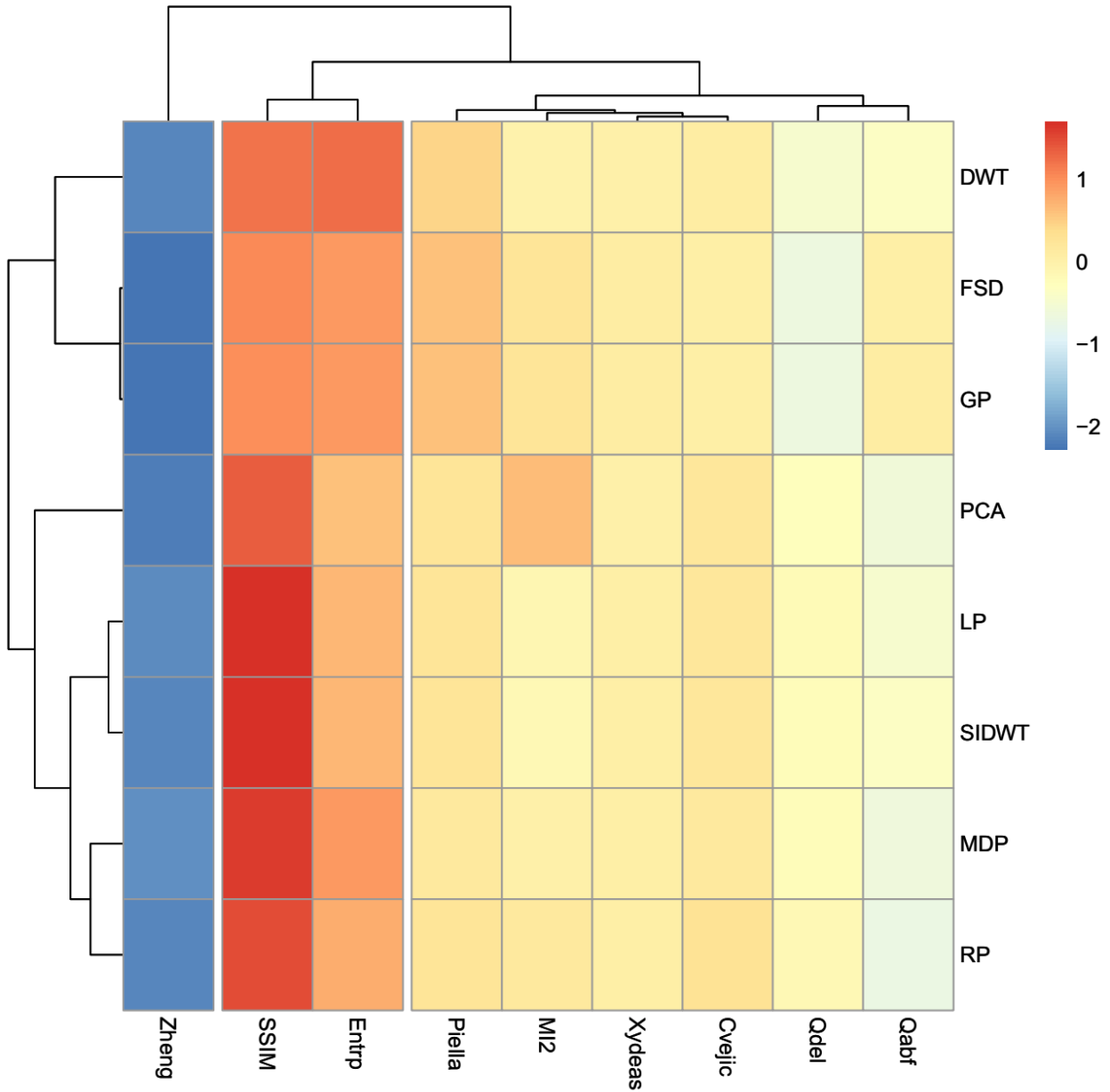

Fig. 12: the pretty heatmap for clustering analysis both metrics and fusion methods in Experiment 8.

### Inference

As can be seen in the ranking table 3, the best metrics that use the information are $Q_{Del}$, Xydas, and Piella. So, in this section, these metrics are focused and analyzed.

At first, the PCA of all 8 experiments on these three metrics are considered. Fig. 13 draws the PCA Biplot of the metrics. The dimensions retained about 97% (93.5% + 3.8%) of the total information contained in the data set. Metrics with similar profile are grouped together. The plot shows how the proposed metric used the information better than the other two metrics; its data are almost closed together in the second quarter of coordinate axes with the average close to the diagonal line.

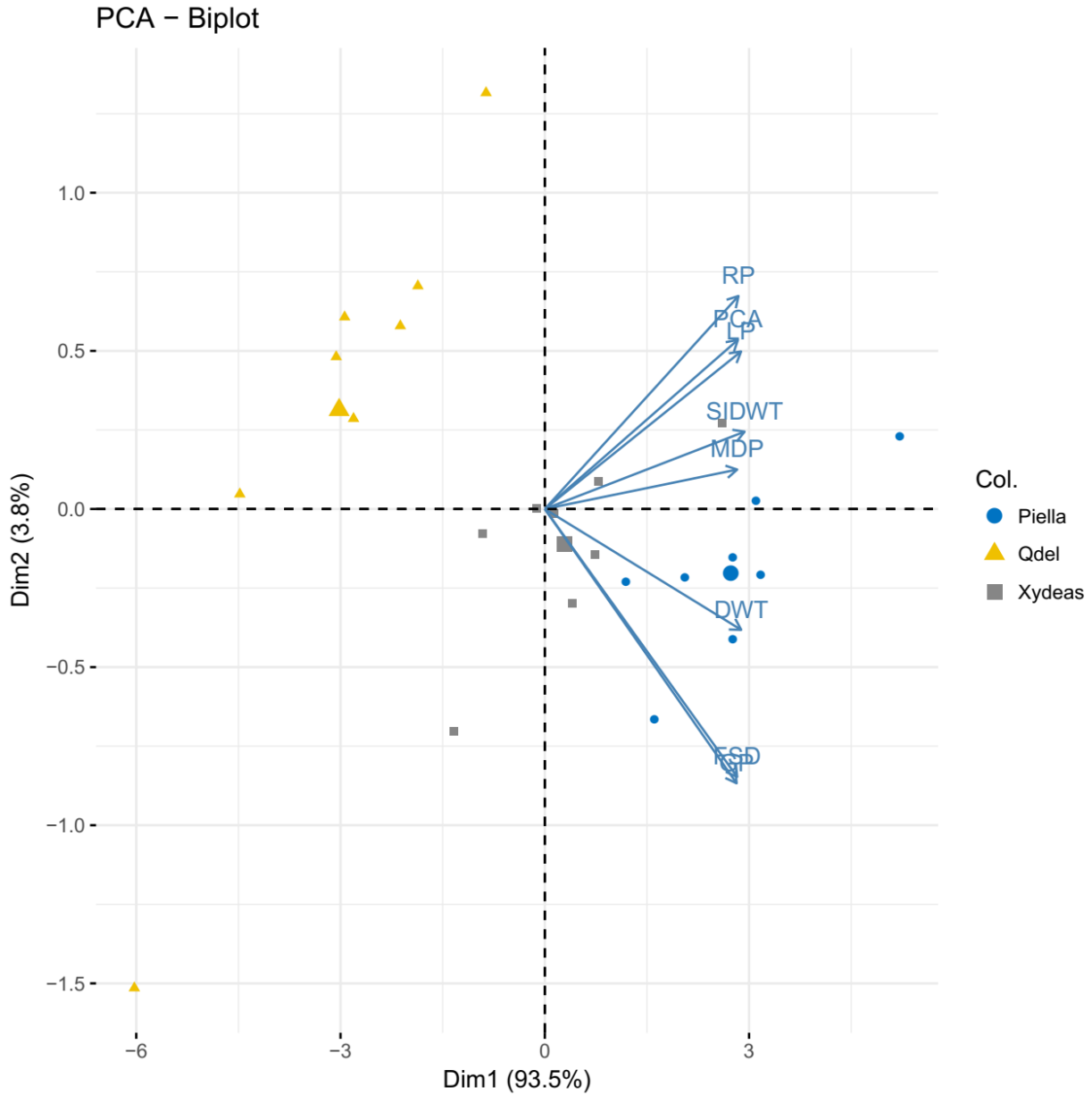

Fig. 13. The PCA-Biplot of three most ranked Objective Evaluation Metrics.

As another analysis, in Fig. 14, the scatter plot matrix of fusion methods are drawn grouped by these three metrics. Each box of the figure is colored blue, yellow and red for the metrics Piella, $Q_{Del}$ and Xydeas, respectively. The figure contains:
- The scatterplots between each pairs of fusion methods (in the lower left boxes of the figure).
- The correlation coefficients, where the black numbers show the correlation of the whole three metrics (in the diagonal mirror of scatterplot boxes).
- The density distributions (on the boxes in diagonal line).
- The box plots in the metrics on the fusion methods (in the last column of the figure).
- The histogram of metrics over the fusion methods (in the last three rows of the figure).

The boxes on the lower left hand side of the whole figure shows the scatterplots of three sets of metrics on the two dimensions of fusion methods. Each metric observation is shown by a filled circle. To show more information, the metrics are shown by its color simultaneously. For example the box in the first column and second row shows data with the horizontal axis of DWT and the vertical axis of FSD.

On the other side of the figure, each upper right box contains the correlation of related fusion methods using the whole three metrics, Piella, $Q_{Del}$ and Xydeas metrics, from top to down, respectively.

The diagonal boxes of the figure plot the distribution of evaluating results of metrics on fusion methods.

Each box in the last column of the figure contains the "box-and-whiskers plot", which is a graphical summary of the metrics on a fusion method. The "hinges" in each box shows the first and third quartiles with a line indicating the median. The "whiskers" display the minimum and maximum values of metrics, in which they fall within a distance of 1.5 times from the closest hinge. The outliers are shown in dots. These information of distribution (such as: min, $1^{st}$ Qu, median, $3^{rd}$ Qu., and Max) are shown side-by-side and can be comparable in a more compact way.

The last three rows of the figure represent the histogram bars of the distribution of metrics. The inputs of these bars are fusion methods and the metrics are cut into several bins. The number of metric observation per bin is shown by the height of the bar.

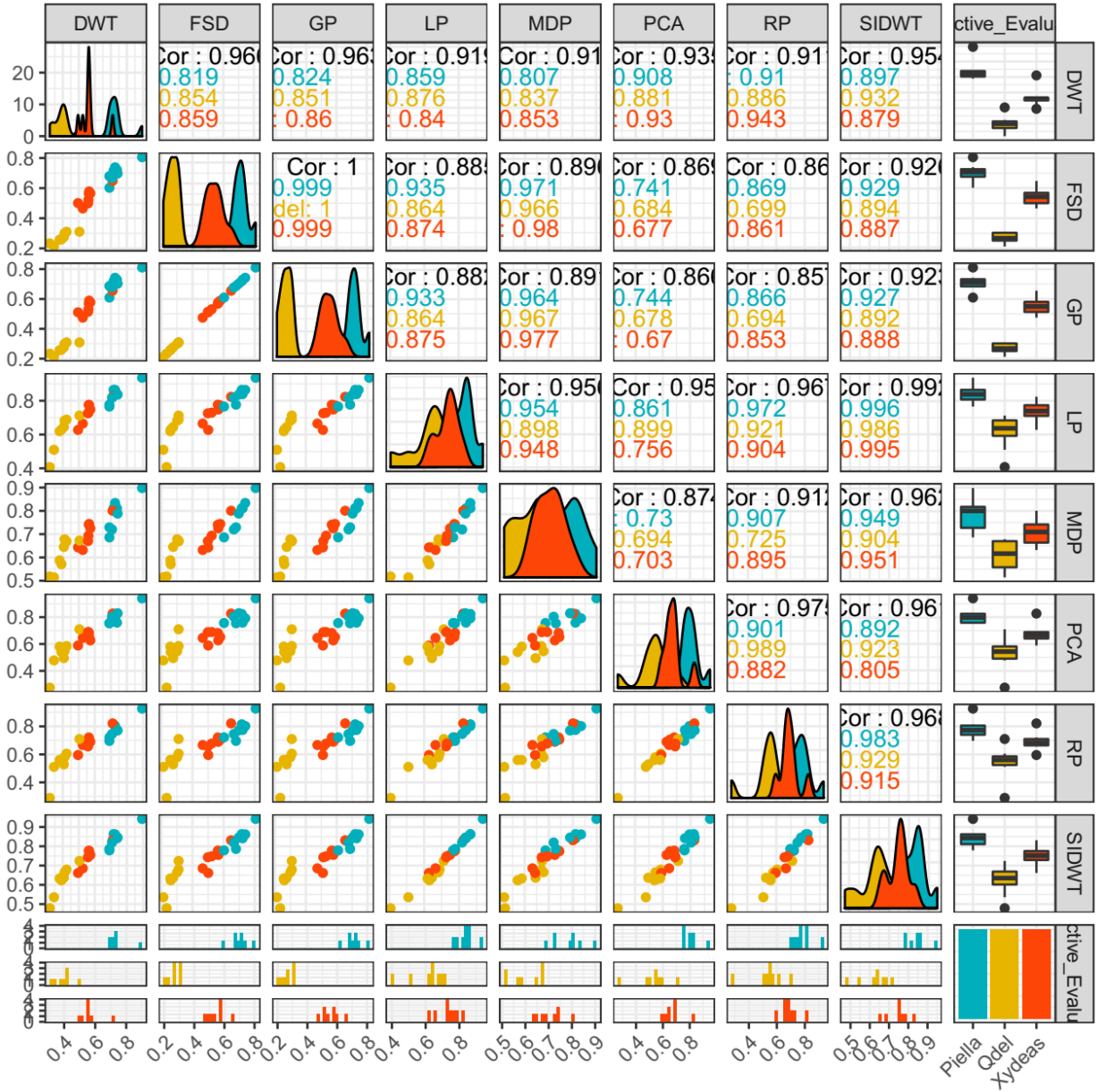

Fig. 14: Scatter plot matrix of fusion methods grouped by three most ranked metrics in Table 4.

## 6. Discussion

In section 4, to statistically judge the quality of the proposed metric, thirteen known objective evaluation metrics were employed on eight fusion methods that fused seven complementary medical images. At first, for each of seven experiments, the fusion methods are evaluated by the metrics. The metrics ranked the fusion methods, experiment by experiment. The average of the ranks was assumed as a more acceptable evaluation metric, in a way that, the smaller average rank a fusion method gained, the better image fusion method is. In the second place, the proposed objective evaluation metric, $Q_{Del}$, was implemented on all source and fused images in seven experiments. At last, the correlation coefficient between each metric's rank and the average of other one's ranks were computed and shown that the proposed metric were more correlated to the average of others in most experiments. Moreover, a general ranking of metric was obtained by averaging ranks of metrics for all experiments. The proposed metric had the best ranked when all experiments were considered.

In this section, the advantages of the proposed objective evaluation metric were discussed.

As the first advantage, $Q_{Del}$ results very close to the informal subjective tests. To illustrate, its results were compared to different known objective evaluation metrics that had been designed to estimate subjective tests. Consequently, the results of $Q_{Del}$ were high correlated to the average decision of those metrics.

The second advantage of $Q_{Del}$ is that, the results are probabilities in the range of [0, 1]. The more $Q_{Del}$ finds a fusion method is powerful, the closer to 1 its result becomes. Actually, the result is the probability that a fusion method has a good performance. So, the proposed metric tends to result close to zero (one) the evaluation of a weak (well) fusion method. Being probability makes the results more understandable and tangible.

Thirdly, $Q_{Del}$ has no parameter. Having no parameter makes the user free of adjusting them case by case. The proposed metric is so dynamic that just gets the source images and the fused one as inputs, and results the probability of how well the source images were fused. On the other hand, two main disadvantages of having parameter in the metric are that regularizing them case by case leads both spent time and change the result.

Fourthly, illumination of the images doesn't affect the result. This is because the new domain makes the images free of illumination dependence. More generally, driving from this property, $Q_{Del}$ even results almost the same for a fused image and its post-processed version by changing the illumination; while they carry the same information. However, there is a little difference between their results; *e.g.* when the contrast is enhanced in the post-processed one, the saliencies become bigger and so, they move a bit in the normal distribution function used in the proposed metric .

Fifthly, the proposed metric concentrates on measuring the transferred information from sources images to the fused one. The information is large saliency of source images and is free of the illumination dependence. The more a pixel in one source image has a large saliency, the more it becomes important in the $Q_{Del}$ evaluation process.

Another advantage is that, $Q_{Del}$ manages the noises. Using Gaussian as a smooth filter makes the effect of noise decreases in the proposed metric. Moreover, averaging all pixels' probability dramatically reduce the effect of remaining infected by noise pixels.

As another advantage, $Q_{Del}$ manages the complementary and conflicting regions. A well fusion method should handle this situation by defining the saliencies on the base of that illumination, where the proposed metric is free of illumination. It easily evaluates the fused image in regions that one source image has large saliencies and another one is pure white or black.

Another advantage is no needing the ground-truth. The proposed metric fetch the requirement information from source images, where compare them with the corresponding information from the fused image.

Another one is that normalizing data make no sense in the proposed metric. $Q_{Del}$ evaluates the fused method based on all source images at once and all together. However, the metrics that mix the evaluation of fused images for each source one should normalize each evaluated result before mixing to be more acceptable metrics; e.g. mixing un-normalized MI leads false result [Hossny et al. 2008].

Demonstrating the quality of visual information is another advantage of $Q_{Del}$. Human visual perception is sensitive to the saliency, where the metric measures the transferred information based on saliency.

Another advantage is that $Q_{Del}$ manages the results when fusion methods are performed on different levels. The multiresolution and non-multiresolution fusion methods try introducing important features and transfer them from the source images to the fused one. At last, those features are acceptable that correlate with the human visual perception. The proposed metric uses a vector domain that is high correlated to the human visual perception, since both of them are sensitive to the differences between neighbor pixels.

Another advantage is that the proposed metric measures the structural similarity. Using Gaussian filter not only decreases the effect of the noise, but also it spreads the information of pixels to their neighbors. While the strong edges have more impact on the neighbors, each pixel is under the influence of its neighbors. Consequently, comparing a pixel in fused image with corresponding pixels in source images means comparing similarity of a region with the center of that pixel.

Having no interest in a fusion method that averages the source images is another advantage of the proposed method. These fusion methods diminish the effect of saliency, and this issue reduces the probabilities made by proposed metric.

We measure the correlation of scores between objective metrics that measure the ranks of same fusion methods. The proposed metric had the most correlation with the average results of previously known metrics, which means it has high equivalence reliability.

Using the new vector field makes the proposed metric to measure the structural similarity and manage the complementary and conflicting regions. These features along with its free of illumination dependence make the proposed metric have high validity.

Moreover, the small time and computational complexity of $Q_{Del}$ are considerable.

## 7. Conclusion

A new objective evaluation metric for image fusion methods, $Q_{Del}$, is presented in this paper. For all source images and the fused one, $Q_{Del}$ first spreads the saliency to the neighbors using Gaussian filter. Then, it transforms the images to a vector domain, using Del operator. The importance of each pixel of source images is determined by the magnitude of these vectors. The corresponding pixels in source images make normal distributions. Then, by converting corresponding pixels in source images and fused one to the standard normal distribution, a probability for each fused pixel is computed. The probabilities demonstrate that how much the pixels in the fused image are better than a set of corresponding pixels created by all fusion methods. Finally, the proposed metric evaluate a fusion method by averaging these probabilities on pixels. To statistically judge the performance of $Q_{Del}$, eight fusion methods that employed on seven complementary medical images, were compared using thirteen well known objective evaluation metrics. The ranks of fusion methods using proposed metric were highly correlated to the average rankings of other metrics. Statistical comparisons consider the proposed metric as the best one among those well-known thirteen. Moreover, according to the analyses carried out in this paper, the proposed metric has the following advantages:

i. The results are very close to the informal subjective tests.
ii. The metric scores any fused image from 0 to 1, which understandable as a probability value. The bigger the value is, the more likely the fusion method fuses well the source images.
iii. It has no parameter. So, the user gets rid of manually regulating parameters case by case.
iv. It is free of illumination dependence.
v. It measures the transferred saliency in the source images.
vi. It measures the structural similarity.
vii. It manages the noise.
viii. It manages the complementary and conflicting regions.
ix. It manages the results when fusion methods employed on different levels.
x. Time and computational complexities are low.
xi. It no longer needs the ground-truth to measure the fusion method's quality.
xii. Normalized data make no sense in this metric.
xiii. It doesn't tend to score the fusion methods that average source images in a way.
xiv. It has high equivalence reliability.
xv. The proposed metric has high validity.

Plus its better results, the vector domain that $Q_{Del}$ uses and the technique of information extraction in this domain cause the proposed objective evaluation metric not only maintains the advantages of previously known metrics, but also gets free from their disadvantages.